\documentclass[twocolumn,showpacs,preprintnumbers,amsmath,amssymb]{revtex4}
\usepackage{pifont}

\usepackage{graphicx}
\usepackage{dcolumn}
\usepackage{bm}

\begin{document}

\preprint{APS/123-QED}

\title{Feynman-Kac Equations for Reaction and Diffusion Processes}

\author{Ru Hou, Weihua Deng}


\affiliation{%
School of Mathematics and Statistics, Gansu Key Laboratory of Applied Mathematics and Complex
Systems, Lanzhou University, Lanzhou 730000, China
\\
}%


\begin{abstract}
This paper provides a theoretical framework of deriving the forward and backward Feynman-Kac equations for the distribution of functionals of the path of a particle undergoing both diffusion and chemical reaction. Very general forms of the equations are obtained. Once given the diffusion type and reaction rate, a specific forward or backward Feynman-Kac equation can be obtained. The listed in the paper include the ones for normal/anomalous diffusions and reactions with linear/nonlinear rates.  Using the derived equations, we also study the occupation time in half-space, the first passage time to a fixed boundary, and the occupation time in half-space with absorbing or reflecting boundary conditions.

\end{abstract}

\maketitle

\section{Introduction}
Diffusion is the net movement of particles from a region of high concentration to a region of low concentration. It is called  normal diffusion if the ensemble mean squared displacement of the diffusing particles scales linearly, otherwise it is called anomalous diffusion, i.e., $\langle x^2(t)\rangle \sim t^\mu$ with $\mu \neq 1$, being subdiffusion for $\mu \in (0,1)$ and superdiffusion for $\mu \in (1,2)$ \cite{Magdziarz2007,Magdziarz2008,Cairoli2015}. For pure diffusion processes (with external forces), the equations governing the distribution of functionals of the path of particle are well established \cite{Cairoli2015a,Cairoli2017}. Influenced by Feynman's thesis of the derivation of Schrodinger's equation, in 1949, Kac derives the classical Feynman-Kac  equation for normal diffusion \cite{Kac1949}. Turgeman et al. present the fractional Feynman-Kac equation for non-Brownian functionals \cite{Turgeman2009} and tempered fractional Feynman-Kac equation for tempered non-Brownian functionals is provided in \cite{Wu2016}.

Chemical reaction leads to the transformation of one type of particles to another type. The main concern of this paper is on the  distribution of the functional of the paths of a particle performing both diffusion and chemical reaction. We first recall the governing equation for the probability density function (PDF) $\rho(x,t)$ of the reaction diffusion process. Assume that the reaction rate law is $\dot{\rho}=r(\rho)\rho$. For normal diffusion, the well-known reaction diffusion equation is
\begin{equation}\label{eq:i01}
  \frac{\partial \rho(x,t)}{\partial t}=K_1\frac{\partial^2\rho(x,t)}{\partial x^2}+r(\rho)\rho,
\end{equation}
with $K_1$ being the diffusion constant. For anomalous diffusion, the situation becomes complicated, and the resulted equation can  not be obtained by simply adding the reaction term into the non-Markovian transport equation. For example, in \cite{Sokolov2006,Henry2006}, the authors derive the subdiffusive reaction diffusion equation:
\begin{equation}\label{eq:i02}
  \frac{\partial \rho(x,t)}{\partial t}=K_\alpha e^{-\kappa t}\mathbb{D}_{t}^{1-\alpha}\left(e^{\kappa t}\frac{\partial^2\rho(x,t)}{\partial x^2}\right)-\kappa \rho(x,t),
\end{equation}
where $K_\alpha$ is the generalized diffusion coefficient and $\mathbb{D}_{t}^{1-\alpha}$ is the Riemann-Liouville fractional derivative. For the multispecies system with anomalous diffusion and reaction dynamics, refer to \cite{Langlands2008,Fedotov2010a,Froemberg2011}, and for the front propagation problems, see \cite{Yadav2007,Iomin2012,Iomin2013}.

Functionals of paths of particles have applications in rather diverse fields, including probability theory, the KPZ varieties, finance, economics, even in characterizing the stochastic behaviour of daily temperature records \cite{Majumdar2005}. So far, it seems there are no research works on the functional distribution of the reaction diffusion processes. Moreover, the functional in all published works is defined as the form $A=\int_0^t U(x(\tau))d\tau$, where $x(t)$ is the particle trajectory and $U(x)$ is a specified function. Here we extend it as $A=\int_0^t U(x(\tau),\tau)d\tau$, which may be used to model the first passage time to a moving boundary \cite{Tuckwell1984}. As for the chemical reactions, both nonlinear reaction rate $r(\rho(x,t))$ and linear reaction rates $r(t)$ and $r(x)$ are considered, i.e., the equations governing reaction process are, respectively, $\dot{\rho}=r(\rho)\rho$, $\dot{\rho}=r(t)\rho$, or $\dot{\rho}=r(x)\rho$. The framework \cite{Turgeman2009} for deriving the Feynman-Kac equations for pure diffusion process does not work for the reaction diffusion process. So, we present new theoretical frameworks of deriving the forward and backward Feynman-Kac equations for the distribution of the more general functionals of the path of a particle undergoing both diffusion and chemical reaction with linear/nonlinear rates; the new framework also applies to the pure diffusion process and the functional $A=\int_0^t U(x(\tau))d\tau$. Based on the very general forms of the derived forward and backward Feynman-Kac equations, we more specifically provide the equations for normal/anomalous diffusions and reactions with linear/nonlinear rate. And the derived equations are used to calculate the occupation time in half-space, the first passage time to a fixed boundary, and the occupation time in half-space with absorbing or reflecting boundary conditions.

The remainder of this paper is organized as follows. In Section \uppercase \expandafter {\romannumeral 2}, we derive the general forward Feynman-Kac equations for the process with  nonlinear reaction rate $r(\rho(x,t))$, and discuss some important properties and examples for the case of linear reaction rates $r(t)$ and $r(x)$. The backward Feynman-Kac equations are presented in Section \uppercase \expandafter {\romannumeral 3}. Following the derived equations, in Section \uppercase \expandafter {\romannumeral 4}, some specific functional distributions are analytically solved, including the occupation time in half-space, the first passage time to a fixed boundary and the occupation time in half-interval with absorbing or reflecting boundary conditions, and some numerical simulations are performed to verify the analytical results. We conclude the paper with some discussions in Section \uppercase \expandafter {\romannumeral 5}.

\section{Derivation of the Forward Feynman-Kac Equations}
Now, we start to derive the forward Feynman-Kac equations for the reaction diffusion processes, composed of three subsections, respectively, for nonlinear reaction rate $r(\rho(x,t))$, linear reaction rate $r(t)$, and linear reaction rate $r(x)$. We limit to the case that all the reaction rates are nonpositive.
\subsection{Nonlinear reaction rate $r(\rho(x,t))$}

 Let $\phi(t)$ and $w(x)$ be the waiting time and symmetric jump length PDF, respectively. We define a stochastic process $\{x(t),A(t)\}$ describing the time-varying positions and functional values that makes physical sense for the reaction diffusion processes. Denote $\rho(x,A,t)$ as the joint PDF of finding the particle at position $x$ with the functional value $A$ at time $t$.  We extend the method in \cite{Fedotov2010} to formulate the balance equations for the density $\rho(x,A,t)$ and the auxiliary density $j(x,A,t)$. The latter is the joint PDF of particles exactly arriving at position $x$ with the functional value $A$ at time $t$. The balance equation for $\rho(x,A,t)$ can be written as
\begin{widetext}
\begin{equation}\label{eq:01}
  \rho(x,A,t)=\rho_{0}(x)\delta\left(A-\int_{0}^{t}U(x,t')dt'\right)e^{\int_{0}^{t}r(\rho(x,u))du}\Psi(t)
    +\int_{0}^{t}j\left(x,A-\int_{\tau}^{t}U(x,t')dt',\tau\right)e^{\int_{\tau}^{t}r(\rho(x,u))du}\Psi(t-\tau)d\tau,
\end{equation}
\end{widetext}
where $\Psi(t)=\int_{t}^{+\infty}\phi(\tau)d\tau$ is the survival probability function, i.e., the probability that the particles keep waiting up to time $t$.
The first term on the right hand side (RHS) of Eq. \eqref{eq:01} represents the particles that  stay at their initial position $x$ up to time $t$. The corresponding functional value of these motionless particles can be calculated as $A=\int_{0}^{t}U(x,t')dt'$. Hence the Dirac delta function $\delta (\cdot)$ is introduced. $r(\rho(x,t))$ denotes the nonlinear reaction rate with $\rho(x,t)=\int_{-\infty}^{+\infty}\rho(x,A,t)d A$. The second term on the RHS of Eq. \eqref{eq:01} describes the particles that arrive at position $x$ at some previous time $\tau<t$ and don't move during the time interval $[\tau,t]$. The difference of the functional values between the time interval $[0,\tau]$ and $[0,t]$ is $\int_{\tau}^{t}U(x,t')dt'$, since no jumps take place during $[\tau,t]$. If the usual functional $A=\int_{0}^{t}U(x(t'))dt'$ is used, one just needs to replace $\int_{0}^{t}U(x,t')dt'$ and $\int_{\tau}^{t}U(x,t')dt'$ in Eq. \eqref{eq:01} with $tU(x)$ and $(t-\tau)U(x)$, respectively.

Moreover, the balance equation for $j(x,A,t)$ can be written in the following form:
\begin{widetext}
\begin{equation}\label{eq:02}
\begin{split}
  j(x,A,t)&=\int_{\mathbb{R}}\rho_{0}(x-z)\delta\left(A-\int_{0}^{t}U(x-z,t')dt'\right)e^{\int_{0}^{t}r(\rho(x-z,u))du}w(z)\phi(t)dz
  \\
  &+\int_{0}^{t}\int_{\mathbb{R}}j\left(x-z,A-\int_{\tau}^{t}U(x-z,t')dt',\tau\right)e^{\int_{\tau}^{t}r(\rho(x-z,u))du}w(z)\phi(t-\tau)dzd\tau.
\end{split}
\end{equation}
\end{widetext}
The first term on the RHS of Eq. \eqref{eq:02} expresses the particles that stay at the initial position $x-z$ and jump to $x$ instantaneously at time $t$. The second term demonstrates the particles that arrive at the position $x-z$ at some previous time $\tau<t$, and react up to time $t$ and jump to $x$ exactly at time $t$.

In order to derive the forward Feynman-Kac equation, we need to conduct the Fourier transforms $A\rightarrow p$ (denoted by $\mathcal{F}\{f(A)\}=\int_{-\infty}^{+\infty}e^{ipA}f(A)dA$) for both Eqs. \eqref{eq:01} and \eqref{eq:02}, since $A$ could be negative.
Then, we obtain
\begin{widetext}
\begin{equation}\label{eq:03}
 \rho(x,p,t)=\rho_{0}(x)e^{ip\int_{0}^{t}U(x,t')dt'}e^{\int_{0}^{t}r(\rho(x,u))du}\Psi(t)
  +\int_{0}^{t}j(x,p,\tau)e^{ip\int_{\tau}^{t}U(x,t')dt'}e^{\int_{\tau}^{t}r(\rho(x,u))du}\Psi(t-\tau)d\tau,
\end{equation}
and
\begin{equation}\label{eq:04}
\begin{split}
j(x,p,t)&=\int_{\mathbb{R}}\rho_{0}(x-z)e^{ip\int_{0}^{t}U(x-z,t')dt'}e^{\int_{0}^{t}r(\rho(x-z,u))du}w(z)\phi(t)dz
\\
&+\int_{0}^{t}\int_{\mathbb{R}}j(x-z,p,\tau)e^{ip\int_{\tau}^{t}U(x-z,t')dt'}e^{\int_{\tau}^{t}r(\rho(x-z,u))du}w(z)\phi(t-\tau)dzd\tau.
\end{split}
\end{equation}

Since Eqs. \eqref{eq:03} and \eqref{eq:04} have nonlinear terms, the standard Fourier-Laplace transform techniques used in \cite{Turgeman2009,Carmi2010,Carmi2011,Wu2016} cannot be applied directly. Instead, we differentiate the density $\rho(x,p,t) $ given by Eq. \eqref{eq:03} with respect to $t$, which leads to

\begin{equation}\label{eq:05}
  \frac{\partial\rho(x,p,t)}{\partial t}=[ipU(x,t)+r(\rho(x,t))]\rho(x,p,t)+j(x,p,t)-i(x,p,t),
\end{equation}
where
\begin{equation}\label{eq:06}
 i(x,p,t)=\rho_{0}(x)e^{ip\int_{0}^{t}U(x,t')dt'}e^{\int_{0}^{t}r(\rho(x,u))du}\phi(t)
  +\int_{0}^{t}j(x,p,\tau)e^{ip\int_{\tau}^{t}U(x,t')dt'}e^{\int_{\tau}^{t}r(\rho(x,u))du}\phi(t-\tau)d\tau.
\end{equation}
\end{widetext}
It follows from Eqs. \eqref{eq:04} and \eqref{eq:06} that
\begin{equation}\label{eq:07}
  j(x,p,t)=\int_{\mathbb{R}}i(x-z,p,t)w(z)dz.
\end{equation}

In order to express $i(x,p,t)$ in terms of $\rho(x,p,t)$, we divide Eqs. \eqref{eq:03} and \eqref{eq:06} by the same factor $e^{ip\int_{0}^{t}U(x,t')dt'}e^{\int_{0}^{t}r(\rho(x,u))du}$ and then take the Laplace transform $t\rightarrow s$ (denoted by $\mathcal{L} \{f(t)\}=\int_{0}^{+\infty}e^{-st}f(t)dt$) for both equations
\begin{widetext}
\begin{equation}\label{eq:08}
  \mathcal{L}\{\rho(x,p,t)e^{-ip\int_{0}^{t}U(x,t')dt'}e^{-\int_{0}^{t}r(\rho(x,u))du}\}=[\rho_{0}(x)+\mathcal{L}\{j(x,p,t)e^{-ip\int_{0}^{t}U(x,t')dt'}e^{-\int_{0}^{t}r(\rho(x,u))du}\}]\tilde{\Psi}(s),
\end{equation}
\begin{equation}\label{eq:09}
  \mathcal{L}\{i(x,p,t)e^{-ip\int_{0}^{t}U(x,t')dt'}e^{-\int_{0}^{t}r(\rho(x,u))du}\}=[\rho_{0}(x)+\mathcal{L}\{j(x,p,t)e^{-ip\int_{0}^{t}U(x,t')dt'}e^{-\int_{0}^{t}r(\rho(x,u))du}\}]\tilde{\phi}(s),
\end{equation}
where $\tilde{\Psi}(s)=\mathcal{L}\{\Psi(t)\}$ and $\tilde{\phi}(s)=\mathcal{L}\{\phi(t)\}$. From Eqs. \eqref{eq:08} and \eqref{eq:09}, we obtain
\begin{equation}\label{eq:10}
 \mathcal{L}\{i(x,p,t)e^{-ip\int_{0}^{t}U(x,t')dt'}e^{-\int_{0}^{t}r(\rho(x,u))du}\}=\frac{\tilde{\phi}(s)}{\tilde{\Psi}(s)}\mathcal{L}\{\rho(x,p,t)e^{-ip\int_{0}^{t}U(x,t')dt'}e^{-\int_{0}^{t}r(\rho(x,u))du}\}. \end{equation}
Performing inverse Laplace transform on Eq. \eqref{eq:10} gives
\begin{equation}\label{eq:11}
 i(x,p,t)=\int_{0}^{t}K(t-\tau)\rho(x,p,\tau)e^{ip\int_{\tau}^{t}U(x,t')dt'}e^{\int_{\tau}^{t}r(\rho(x,u))du}d\tau,
\end{equation}
\end{widetext}
where $K(t)$ is the memory kernel defined by its Laplace transform \cite{Fedotov2010}
\begin{equation}\label{eq:12}
\tilde{K}(s)=\frac{\tilde{\phi}(s)}{\tilde{\Psi}(s)}=\frac{s\tilde{\phi}(s)}{1-\tilde{\phi}(s)}.
\end{equation}
Substituting Eq. \eqref{eq:11} into Eqs. \eqref{eq:05} and \eqref{eq:07} gives the general forward Feynman-Kac equation with nonlinear reaction rate $r(\rho(x,t))$, i.e.,
\begin{widetext}
\begin{equation}\label{eq:13}
\begin{split}
\frac{\partial\rho(x,p,t)}{\partial t}&=\int_{\mathbb{R}}\int_{0}^{t}K(t-\tau)\rho(x-z,p,\tau)e^{ip\int_{\tau}^{t}U(x-z,t')dt'}e^{\int_{\tau}^{t}r(\rho(x-z,u))du}w(z)d\tau dz
\\
&-\int_{0}^{t}K(t-\tau)\rho(x,p,\tau)e^{ip\int_{\tau}^{t}U(x,t')dt'}e^{\int_{\tau}^{t}r(\rho(x,u))du}d\tau+[ipU(x,t)+r(\rho(x,t))]\rho(x,p,t),
\end{split}
\end{equation}
where $K(t)$ is defined by Eq. \eqref{eq:12}. This generalized Feynman-Kac equation has few restrictions on the forms of the waiting time and jump length densities, which allows us to conveniently model standard transport, anomalous diffusion or tempered dynamics together with nonlinear reaction terms.

Obviously, plugging $r(\rho(x,t))\equiv0$ into Eq. \eqref{eq:13} gives the forward Feynman-Kac equation for the pure diffusion process without reactions. To avoid confusion, we denote this equation in terms of $n(x,p,t)$, namely,
\begin{equation}\label{eq:13-1}
\begin{split}
\frac{\partial n(x,p,t)}{\partial t}&=\int_{\mathbb{R}}\int_{0}^{t}K(t-\tau)n(x-z,p,\tau)e^{ip\int_{\tau}^{t}U(x-z,t')dt'}w(z)d\tau dz
\\
&-\int_{0}^{t}K(t-\tau)n(x,p,\tau)e^{ip\int_{\tau}^{t}U(x,t')dt'}d\tau+ipU(x,t)n(x,p,t).
\end{split}
\end{equation}

Assume that the symmetric jump length distribution $w(z)$ satisfies $\int_{\mathbb{R}}z^2w(z)dz=\sigma^2<+\infty$. Expanding the first term on the RHS of Eq. \eqref{eq:13} for small $z$ and truncating the Taylor series at the second order (see Appendix A for more details) gives
\begin{equation}\label{eq:14}
\frac{\partial\rho(x,p,t)}{\partial t}=\frac{\sigma^2}{2}\frac{\partial^2}{\partial x^2}\int_{0}^{t}K(t-\tau)\rho(x,p,\tau)e^{ip\int_{\tau}^{t}U(x,t')dt'}e^{\int_{\tau}^{t}r(\rho(x,u))du}d\tau+[ipU(x,t)+r(\rho(x,t))]\rho(x,p,t).
\end{equation}
\end{widetext}

Now we consider several special cases of Eqs. \eqref{eq:13} and \eqref{eq:14}. To start with, if $p=0$, then $\rho(x,0,t)=\int_{-\infty}^{+\infty}e^{ipA}\rho(x,A,t)dA=\rho(x,t)$. Thus, when $p=0$, Eqs. \eqref{eq:13} and \eqref{eq:14} reduce to Eqs. (14) and (22) in \cite{Fedotov2010} respectively, which describe the nonlinear master equations for reaction diffusion processes. Furthermore, we can specify the waiting time distributions and write down the corresponding Feynman-Kac equations in the continuum limit of large scales and long time as follows.

{(a)} Exponential waiting time distribution

Suppose $\phi(t)=\frac{1}{\tau} \exp(-\frac{t}{\tau})$. Thus, $\tilde{\phi}(s)=\int_{0}^{+\infty}e^{-st}\phi(t)dt=\frac{1}{\tau s+1}$ and $\tilde{K}(s)=\frac{s\tilde{\phi}(s)}{1-\tilde{\phi}(s)}=\frac{1}{\tau} $. Inserting $K(t)=\frac{1}{\tau}\delta(t)$ into Eq. \eqref{eq:14} gives
\begin{equation}\label{eq:15}
\begin{split}
\frac{\partial\rho(x,p,t)}{\partial t}&=\frac{\sigma^2}{2\tau}\frac{\partial^2\rho(x,p,t)}{\partial x^2}+ipU(x,t)\rho(x,p,t)
\\
&+r(\rho(x,t))\rho(x,p,t),
\end{split}
\end{equation}
which is the standard Feynman-Kac equation for the normal diffusion (Markovian) process with the reaction rate $r(\rho(x,t))$.

{(b)} Power-law waiting time distribution

When $\phi(t)=\alpha \tau ^{\alpha} t^{-1-\alpha}\mathbf{1}_{[\tau, +\infty)}(t)$ with $0<\alpha<1$, we can calculate (see Appendix B) $\tilde{\phi}(s)$, which has the asymptotic form $\tilde{\phi}(s)\sim1-\tau ^{\alpha}\Gamma (1-\alpha)s ^{\alpha}$ for $s\rightarrow0$. Meanwhile, according to Eq. \eqref{eq:12}, $\tilde{K}(s)\approx\frac{s^{1-\alpha}}{\tau ^{\alpha}\Gamma(1-\alpha)}$.
Thus, Eq. \eqref{eq:14} reduces to
\begin{widetext}
\begin{equation}\label{eq:16}
\begin{split}
\frac{\partial\rho(x,p,t)}{\partial t}&=\frac{\sigma^2}{2\tau^{\alpha}\Gamma(1-\alpha)}\frac{\partial^2}{\partial x^2}\left[e^{ip\int_{0}^{t}U(x,t')dt'}e^{\int_{0}^{t}r(\rho(x,u))du}\mathbb{D}_{t}^{1-\alpha}\{\rho(x,p,t)e^{-ip\int_{0}^{t}U(x,t')dt'}e^{-\int_{0}^{t}r(\rho(x,u))du}\}\right]
\\
&+[ipU(x,t)+r(\rho(x,t))]\rho(x,p,t),
\end{split}
\end{equation}
where $\mathbb{D}_{t}^{1-\alpha}$ is the Riemann-Liouville fractional derivative defined as $\mathbb{D}_{t}^{1-\alpha}\rho(x,p,t)=\frac{1}{\Gamma(\alpha)}\frac{\partial}{\partial t}\int_{0}^{t}\frac{\rho(x,p,t')}{(t-t')^{1-\alpha}}dt'$.
\end{widetext}

{(c)} Tempered power-law waiting time distribution

Assume $\phi(t)=C_{\tau}^{-1}e^{-\lambda t}t^{-1-\alpha}\mathbf{1}_{[\tau, +\infty)}(t)$, where $0<\alpha<1$ and $C_{\tau}=\int_{\tau}^{+\infty}e^{-\lambda t}t^{-1-\alpha}dt$. Then it can be calculated (see also Appendix B) that $\tilde{\phi}(s)$ has the asymptotic form $\tilde{\phi}(s)\sim1-\frac{\Gamma(1-\alpha)}{\alpha C_{\tau}}(s+\lambda)^{\alpha}+\frac{\Gamma(1-\alpha)}{\alpha C_{\tau}}\lambda^{\alpha}$ for $s\rightarrow0$. Similarly, $\tilde{K}(s)\approx \frac{\alpha C_{\tau}}{\Gamma(1-\alpha)}\frac{s}{(s+\lambda)^{\alpha}-\lambda ^{\alpha}}$.
Under this condition, Eq. \eqref{eq:14} can be rewritten as
\begin{widetext}
\begin{equation}\label{eq:17}
\begin{split}
\frac{\partial\rho(x,p,t)}{\partial t}&=\frac{\sigma^2 \alpha C_{\tau}}{2\Gamma(1-\alpha)}\frac{\partial^2}{\partial x^2}\left[e^{ip\int_{0}^{t}U(x,t')dt'}e^{\int_{0}^{t}r(\rho(x,u))du}\frac{\partial}{\partial t}\int_{0}^{t}H(t-\tau)\rho(x,p,\tau)e^{-ip\int_{0}^{\tau}U(x,t')dt'}e^{-\int_{0}^{\tau}r(\rho(x,u))du}d\tau\right]
\\
&+[ipU(x,t)+r(\rho(x,t))]\rho(x,p,t),
\end{split}
\end{equation}
where $H(t)=e^{-\lambda t}t^{\alpha -1}E_{\alpha,\alpha}(\lambda^{\alpha}t^{\alpha})$ and its Laplace transform $\tilde{H}(s)=\frac{1}{(s+\lambda)^{\alpha}-\lambda ^{\alpha}}$. Here $E_{\alpha,\alpha}(\cdot)$ is the two-parameter Mittag-Leffler function \cite{Podlubny1999}, which is defined as $E_{\alpha,\beta}(z)=\sum\limits_{k=0}^{\infty}\frac{z^{k}}{\Gamma(\alpha k+\beta)}$ with $\alpha>0$ and $\beta>0$ .

\subsection{Linear reaction rate $r(t)$}
Following the similar derivation procedure to the previous subsection, we obtain the forward Feynman-Kac equation with linear reaction rate $r(t)$, i.e.,
\begin{equation}\label{eq:19}
\begin{split}
\frac{\partial\rho(x,p,t)}{\partial t}&=\int_{\mathbb{R}}\int_{0}^{t}K(t-\tau)\rho(x-z,p,\tau)e^{ip\int_{\tau}^{t}U(x-z,t')dt'}e^{\int_{\tau}^{t}r(u)du}w(z)d\tau dz
\\
&-\int_{0}^{t}K(t-\tau)\rho(x,p,\tau)e^{ip\int_{\tau}^{t}U(x,t')dt'}e^{\int_{\tau}^{t}r(u)du}d\tau+[ipU(x,t)+r(t)]\rho(x,p,t),
\end{split}
\end{equation}
where $K(t)$ is defined by Eq. \eqref{eq:12}.
\end{widetext}
By simple calculations, it can be easily checked that $\rho(x,p,t)=n(x,p,t)e^{\int_{0}^{t}r(u)du}$ satisfies Eq. (\ref{eq:19}) if $n(x,p,t)$ is the solution Eq. \eqref{eq:13-1}, which means that for the reaction diffusion process with linear reaction rate $r(t)$, the effect of transport with memory and the linear reaction rate dependening on time $t$ can be decoupled. In fact, this also implies that the conclusion of Eqs. (15-17) in \cite{Fedotov2010} for the case of constant reaction rate still holds for the linear reaction rate, and even for describing the distribution of functionals.


\subsection{Linear reaction rate $r(x)$}
We further consider the linear reaction rate $r(x)$ and the traditional functional definition $A=\int_0^tU(x(\tau))d \tau$, in order to give another derivation of the corresponding forward Feynman-Kac equation. To start with, under these conditions, we have
\begin{widetext}
\begin{equation}\label{eq:c1}
  \rho(x,A,t)=\rho_0(x)\delta(A-tU(x))e^{r(x)t}\Psi(t)+\int_0^tj(x,A-(t-\tau)U(x),\tau)e^{(t-\tau)r(x)}\Psi(t-\tau)d\tau,
\end{equation}
and
\begin{equation}\label{eq:c2}
\begin{split}
  j(x,A,t)&=\int_{\mathbb{R}}\rho_{0}(x-z)\delta(A-tU(x-z))e^{r(x-z)t}w(z)\phi(t)dz
  \\
  &+\int_{0}^{t}\int_{\mathbb{R}}j(x-z,A-(t-\tau)U(x-z),\tau)e^{(t-\tau)r(x-z)}w(z)\phi(t-\tau)dzd\tau.
\end{split}
\end{equation}
Performing Fourier transforms $A\rightarrow p$ on both sides of Eq. \eqref{eq:c1} and Eq. \eqref{eq:c2} gives that
\begin{equation}\label{eq:c3}
  \rho(x,p,t)=\rho_0(x)e^{iptU(x)}e^{r(x)t}\Psi(t)+\int_0^tj(x,p,\tau)e^{ip(t-\tau)U(x)}e^{(t-\tau)r(x)}\Psi(t-\tau)d\tau,
\end{equation}
and
\begin{equation}\label{eq:c4}
\begin{split}
  j(x,p,t)&=\int_{\mathbb{R}}\rho_{0}(x-z)e^{iptU(x-z)}e^{r(x-z)t}w(z)\phi(t)dz
  \\
  &+\int_{0}^{t}\int_{\mathbb{R}}j(x-z,p,\tau)e^{ip(t-\tau)U(x-z)}e^{(t-\tau)r(x-z)}w(z)\phi(t-\tau)dzd\tau.
\end{split}
\end{equation}
Conducting Laplace transform $t\rightarrow s$ to Eqs. \eqref{eq:c3} and \eqref{eq:c4}, respectively, we obtain
\begin{equation}\label{eq:c5}
  \rho(x,p,s)=\rho_0(x)\tilde{\Psi}(s-ipU(x)-r(x))+j(x,p,s)\tilde{\Psi}(s-ipU(x)-r(x)),
\end{equation}
and
\begin{equation}\label{eq:c6}
  j(x,p,s)=\int_{\mathbb{R}}\rho_{0}(x-z)w(z)\tilde{\phi}(s-ipU(x-z)-r(x-z))dz
  +\int_{\mathbb{R}}j(x-z,p,s)w(z)\tilde{\phi}(s-ipU(x-z)-r(x-z))dz.
\end{equation}
Conducting Fourier transforms $x\rightarrow k$ to Eqs. \eqref{eq:c5} and \eqref{eq:c6} gives
\begin{equation}\label{eq:c7}
  \rho(k,p,s)=\tilde{\Psi}\left(s-ip U\left(-i\frac{\partial}{\partial k}\right)-r\left(-i\frac{\partial}{\partial k}\right)\right)[\rho_0(k)+j(k,p,s)],
\end{equation}
and
\begin{equation}\label{eq:c8}
  j(k,p,s)=\hat{w}(k)\tilde{\phi}\left(s-ipU\left(-i\frac{\partial}{\partial k}\right)-r\left(-i\frac{\partial}{\partial k}\right)\right)[\rho_0(k)+j(k,p,s)],
\end{equation}
since we have the identity \cite{Carmi2010} $\mathcal{F}\{xf(x)\}=-i\frac{\partial}{\partial k}\hat{f}(k)$. Substituting $j(k,p,s)=\frac{\hat{w}(k)\tilde{\phi}\left(s-ipU\left(-i\frac{\partial}{\partial k}\right)-r\left(-i\frac{\partial}{\partial k}\right)\right)\rho_0(k)}{1-\hat{w}(k)\tilde{\phi}\left(s-ipU\left(-i\frac{\partial}{\partial k}\right)-r\left(-i\frac{\partial}{\partial k}\right)\right)}$ into Eq. \eqref{eq:c7}, we get
\begin{equation}\label{eq:c9}
  \rho(k,p,s)=\frac{\tilde{\Psi}\left(s-ipU\left(-i\frac{\partial}{\partial k}\right)-r\left(-i\frac{\partial}{\partial k}\right)\right)\rho_0(k)}{1-\hat{w}(k)\tilde{\phi}\left(s-ipU\left(-i\frac{\partial}{\partial k}\right)-r\left(-i\frac{\partial}{\partial k}\right)\right)}.
\end{equation}
\end{widetext}

Furthermore, we specify three kinds of typical waiting time and jump length distributions and then calculate their asymptotic forms in the Laplace or Fourier domain as $s\rightarrow 0$ or $k\rightarrow 0$. Naturally, the inverse Fourier-Laplace transform technique is applied to write down the respective forward Feynman-Kac equations in the long time and large scales limits.

{(a)} Exponential waiting time and Gaussian jump length distributions

Suppose $\phi(t)=\frac{1}{\tau} \exp(-\frac{t}{\tau})$ and $w(x)=\frac{1}{\sqrt{2\pi\sigma^2}}e^{-\frac{x^2}{2\sigma^2}}$. Then $\tilde{\phi}(s)=\frac{1}{1+\tau s}\sim 1-\tau s$ and $\hat{w}(k)=\int_{-\infty}^{+\infty}e^{ikx}w(x)dx=e^{-\frac{1}{2}\sigma^2 k^2}\sim 1-\frac{\sigma^2 k^2}{2}$. Substituting the asymptotic forms of $\tilde{\phi}(s)$ and $\hat{w}(k)$ into Eq. \eqref{eq:c9} and conducting inverse Fourier-Laplace transform, we have
\begin{equation}\label{eq:c10}
  \frac{\partial \rho(x,p,t)}{\partial t}=\frac{\sigma^2}{2\tau}\frac{\partial^2\rho(x,p,t)}{\partial x^2}+[ipU(x)+r(x)]\rho(x,p,t).
\end{equation}

{(b)} Tempered power-law waiting time and Gaussian jump length distributions

Plugging $\tilde{\phi}(s)\sim1-\frac{\Gamma(1-\alpha)}{\alpha C_{\tau}}(s+\lambda)^{\alpha}+\frac{\Gamma(1-\alpha)}{\alpha C_{\tau}}\lambda^{\alpha}$ and $\hat{w}(k)\sim 1-\frac{\sigma^2 k^2}{2}$ into Eq. \eqref{eq:c9} and performing inverse Laplace-Fourier transform give that
\begin{widetext}
\begin{equation}\label{eq:c11}
 \frac{\partial\rho(x,p,t)}{\partial t}=\frac{\alpha C_\tau \sigma^2}{2\Gamma(1-\alpha)}\frac{\partial ^2}{\partial x^2}\left(\frac{\partial}{\partial t}-r(x)-ipU(x)\right)\int_0^t H(t-\tau;x)\rho(x,p,\tau)d\tau+[r(x)+ipU(x)]\rho(x,p,t),
\end{equation}
where $H(t;x)$ is defined by its Laplace transform $\tilde{H}(s;x)=\frac{1}{(s+\lambda-r(x)-ipU(x))^\alpha-\lambda^\alpha}$. It can also be written as
\begin{equation}\label{eq:c12}
 \frac{\partial\rho(x,p,t)}{\partial t}=\frac{\alpha C_\tau \sigma^2}{2\Gamma(1-\alpha)}\frac{\partial ^2}{\partial x^2}\mathbb{D}_{t}^{1-\alpha,\lambda,x}\rho(x,p,t)+[\lambda^\alpha\mathbb{D}_{t}^{1-\alpha,\lambda,x}-\lambda][\rho(x,p,t)-e^{iptU(x)}e^{r(x)t}\rho_0(x)]+[r(x)+ipU(x)]\rho(x,p,t),
\end{equation}
where the tempered fractional substantial derivative $\mathbb{D}_{t}^{1-\alpha,\lambda,x}$ is defined in the Laplace domain as $\mathcal{L} \{\mathbb{D}_{t}^{1-\alpha,\lambda,x}\rho(x,p,t)\}=(s+\lambda-r(x)-ipU(x))^{1-\alpha}\rho(x,p,s)$. And in the time domain,
\begin{equation}\label{eq:c13}
 \mathbb{D}_{t}^{1-\alpha,\lambda,x}\rho(x,p,t)=\frac{1}{\Gamma(\alpha)}\left[\frac{\partial}{\partial t}+\lambda-r(x)-ipU(x)\right]\int_0^t\frac{e^{(t-\tau)(r(x)+ipU(x)-\lambda)}}{(t-\tau)^{1-\alpha}}\rho(x,p,\tau)d\tau,
 \end{equation}
which is equivalent to
\begin{equation}\label{eq:c14}
  \mathbb{D}_{t}^{1-\alpha,\lambda,x}\rho(x,p,t)=\frac{e^{iptU(x)+r(x)t-\lambda t}}{\Gamma(\alpha)}\frac{\partial}{\partial t}\int_0^t\frac{e^{\lambda \tau-ip\tau U(x)-r(x)\tau}\rho(x,p,\tau)}{(t-\tau)^{1-\alpha}}d\tau.
\end{equation}
\end{widetext}
It should be noted that Eqs. \eqref{eq:c11} and \eqref{eq:c12} are the generalizations of Eqs. (11) and (5) in \cite{Wu2016} respectively to the reaction diffusion cases.

{ (c)} Tempered power-law waiting time and jump length distributions

In this case, we assume that the waiting time and jump length obey different tempered power-law distributions. Let $w(x)=C_{\varepsilon}^{-1}e^{-\gamma |x|}|x|^{-1-\beta}\mathbf{1}_{[\varepsilon, +\infty)}(|x|)$ for $0<\beta<2$. The normalization factor $C_{\varepsilon}$ is defined as $C_{\varepsilon}=\int_{\varepsilon}^{+\infty}e^{-\gamma x}x^{-1-\beta}dx+\int_{-\infty}^{-\varepsilon}e^{-\gamma |x|}|x|^{-1-\beta}dx$ to make sure $\int_{-\infty}^{+\infty}w(x)dx=1$. Then it could be calculated (see Appendix C) that $\hat{w}(k)$ has the asymptotic form $\hat{w}(k)\sim1-\frac{2\Gamma(1-\beta)}{\beta C_{\varepsilon}}(k^2+\gamma^2)^{\beta/2}+\frac{2\Gamma(1-\beta)}{\beta C_{\varepsilon}}\gamma^{\beta}$ for $k \rightarrow 0$. Substituting the asymptotic $\tilde{\phi}(s)$ and $\hat{w}(k)$ into Eq. \eqref{eq:c9} and conducting inverse Laplace-Fourier transform, we obtain
\begin{widetext}
\begin{equation}\label{eq:c15}
  \frac{\partial\rho(x,p,t)}{\partial t}=\frac{2\alpha C_\tau \Gamma(1-\beta)}{\beta C_{\varepsilon}\Gamma(1-\alpha)}(\nabla_x^{\beta,\gamma}+\gamma^{\beta})\left(\frac{\partial}{\partial t}-r(x)-ipU(x)\right)\int_0^t H(t-\tau;x)\rho(x,p,\tau)d\tau+[r(x)+ipU(x)]\rho(x,p,t),
\end{equation}
where the tempered fractional Riesz derivative $\nabla_x^{\beta,\gamma}$ (see \cite{Wu2016} for more details) is defined in the Fourier domain as $\mathcal{F}\{\nabla_x^{\beta,\gamma}\rho(x,p,t)\}=-(k^2+\gamma^2)^{\beta/2}\rho(k,p,t)$. Equivalently,
\begin{equation}\label{eq:c16}
\begin{split}
 \frac{\partial\rho(x,p,t)}{\partial t}&=\frac{2\alpha C_\tau \Gamma(1-\beta)}{\beta C_{\varepsilon}\Gamma(1-\alpha)}(\nabla_x^{\beta,\gamma}+\gamma^{\beta})\mathbb{D}_{t}^{1-\alpha,\lambda,x}\rho(x,p,t)+[\lambda^\alpha\mathbb{D}_{t}^{1-\alpha,\lambda,x}-\lambda][\rho(x,p,t)-e^{iptU(x)}e^{r(x)t}\rho_0(x)]
 \\
 &+[r(x)+ipU(x)]\rho(x,p,t),
\end{split}
\end{equation}
where the operator $\mathbb{D}_{t}^{1-\alpha,\lambda,x}$ is defined as Eq. \eqref{eq:c13} or Eq. \eqref{eq:c14}. When $r(x)\equiv0$, Eq. \eqref{eq:c16} reduces to Eq. (36) in \cite{Wu2016}.

\section{Derivation of the Backward Feynman-Kac Equations}
Let $\rho_{x_0}(A,t)$ be the PDF of the functional $A$ at time $t$ with the initial position $x_0$. The backward Feynman-Kac equation regarding $\rho_{x_0}(A,t)$ with the nonlinear reaction rate $r(\rho(x,t))$ can be written as follows
\begin{equation}\label{eq:20}
\begin{split}
 \rho_{x_0}(A,t)&=\delta\left(A-\int_{0}^{t}U(x_0,t')dt'\right)e^{\int_{0}^{t}r(\rho(x_0,u))du}\Psi(t)
 \\
 &+\int_{\mathbb{R}}\int_0^{t}w(z)\phi(\tau)\rho_{x_0+z}\left(A-\int_0^{\tau}U(x_0,t')dt',t-\tau\right)e^{\int_{0}^{\tau }r(\rho(x_0,u))du}d\tau dz.
\end{split}
\end{equation}
\end{widetext}
It should be noted that $\rho(x_0,t)\neq \int_{-\infty}^{+\infty}\rho_{x_0}(A,t)d A$. While the first term on the RHS of Eq. \eqref{eq:20} indicates the motionless particles remaining at their initial position $x_0$ up to time $t$, the second term alternatively represents the particles that jump to the location $x_0+z$ at time $\tau$ ($\tau<t$). Similarly, conducting the Fourier transform $A\rightarrow p$, we obtain
\begin{widetext}
\begin{equation}\label{eq:21}
 \rho_{x_0}(p,t)=e^{ip\int_0^tU(x_0,t')dt'}e^{\int_{0}^{t}r(\rho(x_0,u))du}\Psi(t)
 +\int_{\mathbb{R}}\int_0^{t}w(z)\phi(\tau)\rho_{x_0+z}(p,t-\tau)e^{ip\int_0^\tau U(x_0,t')dt'}e^{\int_{0}^{\tau }r(\rho(x_0,u))du}d\tau dz.
\end{equation}

In what follows, we consider a special case of Eq. \eqref{eq:21}, which supposes the reaction rate is $r(x)$ and the functional is defined as $A=\int_0^t U[x(t')]dt'$. Thus, under these assumptions, Eq. \eqref{eq:21} can be rewritten as
\begin{equation}\label{eq:22}
 \rho_{x_0}(p,t)=e^{iptU(x_0)}e^{r(x_0)t}\Psi(t)
 +\int_{\mathbb{R}}\int_0^{t}w(z)\phi(\tau)\rho_{x_0+z}(p,t-\tau)e^{ip\tau U(x_0)}e^{r(x_0)\tau}d\tau dz.
\end{equation}
\end{widetext}
Consequently, the standard Laplace-Fourier transform technique is applicable to further simplify the concerned equation. Conducting to Eq. \eqref{eq:22} the Laplace transform $t\rightarrow s$ and the Fourier transform $x_0\rightarrow k$, we get
\begin{equation}\label{eq:23}
\rho_{k}(p,s)=\frac{\tilde{\Psi}\left(s-r\left(-i\frac{\partial}{\partial k}\right)-ipU\left(-i\frac{\partial}{\partial k}\right)\right)\delta(k)}{1-\tilde{\phi}\left(s-r\left(-i\frac{\partial}{\partial k}\right)-ipU\left(-i\frac{\partial}{\partial k}\right)\right)\hat{w}(k)}.
\end{equation}
Here we list three kinds of typical waiting time and jump length distributions and write down the respective backward equations in the continuum limit. We omit the power-law waiting time distribution cases, since they could be recovered when the exponential tempering exponents are set to zeros.

{(a)} Exponential waiting time and Gaussian jump length distributions

 Plugging $\tilde{\phi}(s)\sim 1-\tau s$ and $\hat{w}(k)\sim 1-\frac{\sigma^2 k^2}{2}$ into Eq. \eqref{eq:23} and performing inverse Laplace-Fourier transform give that
\begin{equation}\label{eq:24}
  \frac{\partial \rho_{x_0}(p,t)}{\partial t}=\frac{\sigma^2}{2\tau}\frac{\partial^2\rho_{x_0}(p,t)}{\partial x_0^2}+[r(x_0)+ipU(x_0)]\rho_{x_0}(p,t).
\end{equation}

{ (b)} Tempered power-law waiting time and Gaussian jump length distributions

With the substitution of $\tilde{\phi}(s)\sim1-\frac{\Gamma(1-\alpha)}{\alpha C_{\tau}}(s+\lambda)^{\alpha}+\frac{\Gamma(1-\alpha)}{\alpha C_{\tau}}\lambda^{\alpha}$ and $\hat{w}(k)\sim 1-\frac{\sigma^2 k^2}{2}$ into Eq. \eqref{eq:23} and inverse Laplace-Fourier transform, we obtain
\begin{widetext}
\begin{equation}\label{eq:25}
 \frac{\partial\rho_{x_0}(p,t)}{\partial t}=\frac{\alpha C_\tau \sigma^2}{2\Gamma(1-\alpha)}\left(\frac{\partial}{\partial t}-r(x_0)-ipU(x_0)\right)\int_0^t H(t-\tau;x_0)\frac{\partial^2 \rho_{x_0}(p,\tau)}{\partial x_0^2}d\tau+[r(x_0)+ipU(x_0)]\rho_{x_0}(p,t),
\end{equation}
where $H(t;x_0)$ is defined by its Laplace transform $\tilde{H}(s;x_0)=\frac{1}{(s+\lambda-r(x_0)-ipU(x_0))^\alpha-\lambda^\alpha}$. Equivalently,
\begin{equation}\label{eq:26}
 \frac{\partial\rho_{x_0}(p,t)}{\partial t}=\frac{\alpha C_\tau \sigma^2}{2\Gamma(1-\alpha)}\mathbb{D}_{t}^{1-\alpha,\lambda,x_0}\frac{\partial^2 \rho_{x_0}(p,t)}{\partial x_0^2}+[\lambda^\alpha\mathbb{D}_{t}^{1-\alpha,\lambda,x_0}-\lambda][\rho_{x_0}(p,t)-e^{iptU(x_0)}e^{r(x_0)t}]+[r(x_0)+ipU(x_0)]\rho_{x_0}(p,t),
\end{equation}
where the tempered fractional substantial derivative $\mathbb{D}_{t}^{1-\alpha,\lambda,x_0}$ is defined in the Laplace domain as $\mathcal{L} \{\mathbb{D}_{t}^{1-\alpha,\lambda,x_0}\rho_{x_0}(p,t)\}=(s+\lambda-r(x_0)-ipU(x_0))^{1-\alpha}\rho_{x_0}(p,s)$. And in the time domain,
\begin{equation}\label{eq:27}
 \mathbb{D}_{t}^{1-\alpha,\lambda,x_0}\rho_{x_0}(p,t)=\frac{1}{\Gamma(\alpha)}\left[\frac{\partial}{\partial t}+\lambda-r(x_0)-ipU(x_0)\right]\int_0^t\frac{e^{(t-\tau)(r(x_0)+ipU(x_0)-\lambda)}}{(t-\tau)^{1-\alpha}}\rho_{x_0}(p,\tau)d\tau,
 \end{equation}
which is equivalent to
\begin{equation}\label{eq:28}
  \mathbb{D}_{t}^{1-\alpha,\lambda,x_0}\rho_{x_0}(p,t)=\frac{e^{iptU(x_0)+r(x_0)t-\lambda t}}{\Gamma(\alpha)}\frac{\partial}{\partial t}\int_0^t\frac{e^{\lambda \tau-ip\tau U(x_0)-r(x_0)\tau}\rho_{x_0}(p,\tau)}{(t-\tau)^{1-\alpha}}d\tau.
\end{equation}

{ (c)} Tempered power-law waiting time and jump length distributions

Substituting $\tilde{\phi}(s)\sim1-\frac{\Gamma(1-\alpha)}{\alpha C_{\tau}}(s+\lambda)^{\alpha}+\frac{\Gamma(1-\alpha)}{\alpha C_{\tau}}\lambda^{\alpha}$ and $\hat{w}(k)\sim1-\frac{2\Gamma(1-\beta)}{\beta C_{\varepsilon}}(k^2+\gamma^2)^{\beta/2}+\frac{2\Gamma(1-\beta)}{\beta C_{\varepsilon}}\gamma^{\beta} $ into Eq. \eqref{eq:23} and doing Laplace-Fourier transform lead to
\begin{equation}\label{eq:29}
 \frac{\partial\rho_{x_0}(p,t)}{\partial t}=\frac{2\alpha C_\tau \Gamma(1-\beta)}{\beta C_{\varepsilon}\Gamma(1-\alpha)}\left(\frac{\partial}{\partial t}-r(x_0)-ipU(x_0)\right)\int_0^t H(t-\tau;x_0)(\nabla_{x_0}^{\beta,\gamma}+\gamma^{\beta})\rho_{x_0}(p,\tau)d\tau+[r(x_0)+ipU(x_0)]\rho_{x_0}(p,t),
\end{equation}
where the operator $\nabla_{x_0}^{\beta,\gamma}$ is defined in the Fourier domain as $\mathcal{F}\{\nabla_{x_0}^{\beta,\gamma}\rho_{x_0}(p,t)\}=-(k^2+\gamma^2)^{\beta/2}\rho_{k}(p,t)$. Equivalently,
\begin{equation}\label{eq:30}
\begin{split}
 \frac{\partial\rho_{x_0}(p,t)}{\partial t}&=\frac{2\alpha C_\tau \Gamma(1-\beta)}{\beta C_{\varepsilon}\Gamma(1-\alpha)}\mathbb{D}_{t}^{1-\alpha,\lambda,x_0}(\nabla_{x_0} ^{\beta,\gamma}+\gamma^{\beta})\rho_{x_0}(p,t)+[\lambda^\alpha\mathbb{D}_{t}^{1-\alpha,\lambda,x_0}-\lambda][\rho_{x_0}(p,t)-e^{iptU(x_0)}e^{r(x_0)t}]
 \\
 &+[r(x_0)+ipU(x_0)]\rho_{x_0}(p,t),
\end{split}
\end{equation}
where the operator $\mathbb{D}_{t}^{1-\alpha,\lambda,x_0}$ is defined as Eq. \eqref{eq:27} or Eq. \eqref{eq:28}.
\end{widetext}

\section{Applications of the derived equations}
In this section, we present the distributions of specific functionals of the paths of particles performing temporal tempered anomalous dynamics with piecewise constant reaction rate. We analytically solve the corresponding backward Feynman-Kac equations for obtaining the distributions, the moments, and other properties of interest. As pointed out in \cite{Wu2016}, our analysis is mainly based on the derived backward Feynman-Kac equations, since here we are more concerned with the functional distributions than the particles' positions. From this perspective, backward equations are more convenient, although the forward equations could lead to the same conclusions with the extra integration over $x$ step. In what follows, we assume the reaction rate function $r(x)$ satisfying
\begin{equation}\label{eq:31}
  r(x)= \left\{
\begin{array}{ll}
\kappa_1, ~~~~~& x> 0,
\\
\kappa_2, ~~~~~& x< 0,
\end{array}
\right.
\end{equation}
where $\kappa_1$ and $\kappa_2$ are negative constants. Here we confine our analysis to negative reaction rates in order to model the spontaneous evanescent process in which the particles are destroyed or removed at different constant rates depending on their positions. As for the reproduction process with positive reaction rates, one must specify the rules regarding the waiting time of the newborn particles \cite{Abad2010,Fedotov2010}, which is a problem to be explored further.

\subsection{Occupation time in half-space and its fluctuations}
Define the occupation time of a particle in the positive half-space as $T^+=\int_0^tU(x(\tau))d\tau$, where $U(x)=1$ for $x\geq0$ and $U(x)=0$ for $x<0$. Since obviously $T^+\geq 0$, we rely on the Laplace transform $T^+\rightarrow p$, instead of the Fourier transform. In order to find the PDF of $T^+$, we consider the backward equation Eq. \eqref{eq:25} (or Eq. \eqref{eq:26}), conduct Laplace transform $t\rightarrow s$, and substitute the assumed $U(x_0)$ and $r(x_0)$. Denote $K_{\alpha}=\frac{\alpha C_\tau \sigma^2}{2\Gamma(1-\alpha)}$. Thus,
\begin{equation}\label{eq:32}
\small \rho_{x_0}(p,s)= \left\{
\begin{array}{ll}
\frac{K_{\alpha}}{(s+\lambda+p-\kappa_1)^{\alpha}-\lambda^{\alpha}}\frac{\partial^2 \rho_{x_0}(p,s)}{\partial x_0^2}+\frac{1}{s+p-\kappa_1}, & x_0> 0;
\\
\frac{K_{\alpha}}{(s+\lambda-\kappa_2)^{\alpha}-\lambda^{\alpha}}\frac{\partial^2 \rho_{x_0}(p,s)}{\partial x_0^2}+\frac{1}{s-\kappa_2}, & x_0< 0.
\end{array}
\right.
\end{equation}

Solving the equations in each half-space individually and demanding $\rho_{x_0}(p,s)<\infty$ for $|x_0|\rightarrow \infty$, we have
\begin{equation}\label{eq:33}
  \rho_{x_0}(p,s)= \left\{
\begin{array}{ll}
C_1e^{-x_0\sqrt{\frac{(s+\lambda+p-\kappa_1)^{\alpha}-\lambda^{\alpha}}{K_{\alpha}}}}+\frac{1}{s+p-\kappa_1}, & x_0> 0;
\\
C_2e^{x_0\sqrt{\frac{(s+\lambda-\kappa_2)^{\alpha}-\lambda^{\alpha}}{K_{\alpha}}}}+\frac{1}{s-\kappa_2}, & x_0< 0.
\end{array}
\right.
\end{equation}
To determine the constants $C_1$ and $C_2$, we require that $\rho_{x_0}(p,s)$ and its first derivative $\frac{\partial \rho_{x_0}(p,s)}{\partial x_0} $ are continuous at $x_0=0$. Consequently, we have
\begin{equation}\label{eq:34}
\left\{
\begin{array}{ll}
C_1=\frac{-(\kappa_1-\kappa_2-p)\sqrt{(s+\lambda-\kappa_2)^{\alpha}-\lambda^{\alpha}}}{(s+p-\kappa_1)(s-\kappa_2)
(\sqrt{(s+\lambda+p-\kappa_1)^{\alpha}-\lambda^{\alpha}}+\sqrt{(s+\lambda-\kappa_2)^{\alpha}-\lambda^{\alpha}})};
\\
C_2=\frac{(\kappa_1-\kappa_2-p)\sqrt{(s+\lambda+p-\kappa_1)^{\alpha}-\lambda^{\alpha}}}{(s+p-\kappa_1)(s-\kappa_2)
(\sqrt{(s+\lambda+p-\kappa_1)^{\alpha}-\lambda^{\alpha}}+\sqrt{(s+\lambda-\kappa_2)^{\alpha}-\lambda^{\alpha}})}.
\end{array}
\right.
\end{equation}

Suppose the particle starts from $x_0=0$. Then
\begin{widetext}
\begin{equation}\label{eq:35}
 \rho_{0}(p,s)=\frac{(s-\kappa_2)\sqrt{(s+\lambda+p-\kappa_1)^{\alpha}-\lambda^{\alpha}}+(s+p-\kappa_1)\sqrt{(s+\lambda-\kappa_2)^{\alpha}-\lambda^{\alpha}}}{(s+p-\kappa_1)(s-\kappa_2)
(\sqrt{(s+\lambda+p-\kappa_1)^{\alpha}-\lambda^{\alpha}}+\sqrt{(s+\lambda-\kappa_2)^{\alpha}-\lambda^{\alpha}})},
\end{equation}
\end{widetext}
which describes the PDF of $T^+$ in the Laplace domain ($T^+\rightarrow p$ and $t\rightarrow s$). When both $\kappa_1$ and $\kappa_2$ are set to zeros in Eq. \eqref{eq:35}, the Eq. (47) in \cite{Wu2016} is recovered, describing the distribution of $T^{+}$ for tempered anomalous motions without reactions. However, it seems difficult to invert Eq. \eqref{eq:35} analytically, even for the $\kappa_1=\kappa_2=0$ case as reported in \cite{Wu2016}, while the occupation fraction $T^{+}/t$ obeys Lamperti distribution for anomalous motions without exponential tempering ($\lambda=0$) and reactions ($\kappa_1=\kappa_2=0$) \cite{Carmi2010}. Specially, when $\alpha=1$, $\rho_{0}(p,s)=(s-\kappa_2)^{-1/2}(s+p-\kappa_1)^{-1/2}$, apparently different from the arcsine law of the occupation fraction for Brownian motion \cite{Majumdar2005,Barkai2006}.

\begin{figure}
\centering
\includegraphics[height=4.0cm,width=8.1cm]{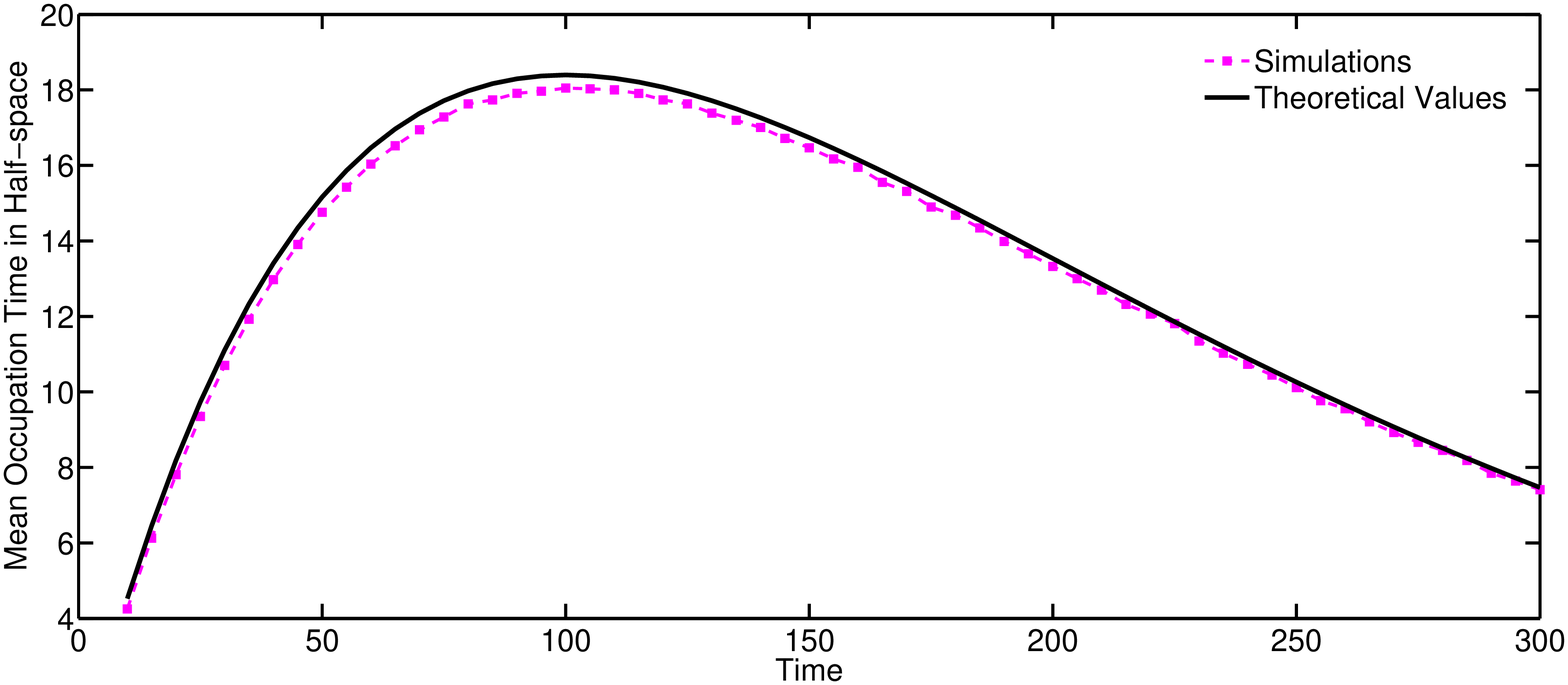}
\caption{$\langle T^{+}\rangle$ simulated from $10^6$ trajectories when $\kappa_1=\kappa_2=-0.01$, $\alpha=0.5$ and $\sigma=1$. The solid (black) line is the derived theoretical evolution of $\langle T^{+}\rangle$, namely $\frac{te^{-0.01 t}}{2}$.}\label{fig1}
\end{figure}

\begin{figure}
\centering
\includegraphics[height=4.0cm,width=8.1cm]{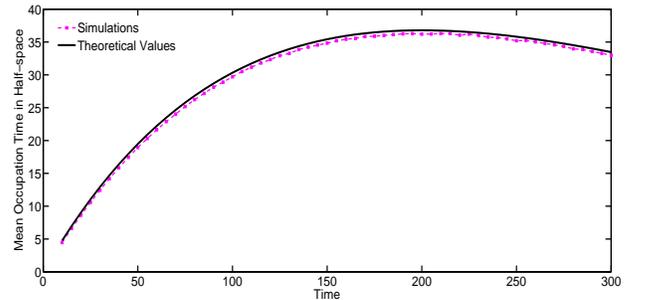}
\caption{$\langle T^{+}\rangle$ simulated from $10^6$ trajectories when $\kappa_1=\kappa_2=-0.005$, $\alpha=0.5$ and $\sigma=1$. The solid (black) line is the derived theoretical evolution of $\langle T^{+}\rangle$, namely $\frac{te^{-0.005 t}}{2}$.}\label{fig2}
\end{figure}

Furthermore, in order to evaluate the expectation and fluctuation of the occupation time, we calculate the moments of $T^+$ in the Laplace domain ($t\rightarrow s$) from Eq. \eqref{eq:35} as follows:
\begin{widetext}
\begin{equation}\label{eq:35-1}
\begin{split}
  \langle T^{+}\rangle_{s}=\left.-\frac{\partial \rho_{0}(p,s)}{\partial p}\right|_{p=0}
    &=\frac{\sqrt{(s+\lambda-\kappa_1)^{\alpha}-\lambda^{\alpha}}}{(s-\kappa_1)^2(\sqrt{(s+\lambda-\kappa_1)^{\alpha}-\lambda^{\alpha}}
  +\sqrt{(s+\lambda-\kappa_2)^{\alpha}-\lambda^{\alpha}})}
  \\
  &+\frac{\alpha(\kappa_2-\kappa_1)(s+\lambda-\kappa_1)^{\alpha-1}\sqrt{(s+\lambda-\kappa_2)^{\alpha}-\lambda^{\alpha}}}
  {2(s-\kappa_1)(s-\kappa_2)\sqrt{(s+\lambda-\kappa_1)^{\alpha}-\lambda^{\alpha}}(\sqrt{(s+\lambda-\kappa_1)^{\alpha}-\lambda^{\alpha}}+\sqrt{(s+\lambda-\kappa_2)^{\alpha}-\lambda^{\alpha}})^2}.
\end{split}
\end{equation}
\end{widetext}
According to Eq. \eqref{eq:35-1}, when $\kappa_1=\kappa_2$, $\langle T^{+}\rangle_s=\frac{1}{2(s-\kappa_1)^2}$, and correspondingly in the time domain $\langle T^{+}\rangle=\frac{te^{\kappa_1 t}}{2}$, or $\langle \frac{T^{+}}{t}\rangle=\frac{e^{\kappa_1 t}}{2}$. Here $e^{\kappa_1 t}$ can be interpreted as the survival probability. This result is further confirmed by the simulation results in Figs. \ref{fig1}-\ref{fig2} and it is remarkably different from the previous studies, for example \cite{Wu2016,CarmiBarkai2011}, which presents $\langle T^{+}\rangle=\frac{t}{2}$, or $\langle \frac{T^{+}}{t}\rangle=\frac{1}{2}$ for (tempered) anomalous diffusion processes.
\begin{widetext}
\begin{equation}\label{eq:35-2}
\begin{split}
 \langle (T^{+})^2\rangle_{s}&=\left.\frac{\partial^2 \rho_{0}(p,s)}{\partial p^2}\right|_{p=0}=\frac{2\sqrt{(s+\lambda-\kappa_1)^{\alpha}-\lambda^{\alpha}}}{(s-\kappa_1)^3(\sqrt{(s+\lambda-\kappa_1)^{\alpha}-\lambda^{\alpha}}
  +\sqrt{(s+\lambda-\kappa_2)^{\alpha}-\lambda^{\alpha}})}
  \\
  &-\frac{\alpha(s+\lambda-\kappa_1)^{\alpha-1}\sqrt{(s+\lambda-\kappa_2)^{\alpha}-\lambda^{\alpha}}}
  {(s-\kappa_1)^2\sqrt{(s+\lambda-\kappa_1)^{\alpha}-\lambda^{\alpha}}(\sqrt{(s+\lambda-\kappa_1)^{\alpha}-\lambda^{\alpha}}+\sqrt{(s+\lambda-\kappa_2)^{\alpha}-\lambda^{\alpha}})^2}
  \\
 &+\frac{\alpha(\kappa_2-\kappa_1)(s+\lambda-\kappa_1)^{\alpha-2} \sqrt{(s+\lambda-\kappa_2)^{\alpha}-\lambda^{\alpha}}\left[B_1(s) \sqrt{(s+\lambda-\kappa_1)^{\alpha}-\lambda^{\alpha}}+B_2(s)\sqrt{(s+\lambda-\kappa_2)^{\alpha}-\lambda^{\alpha}}\right]}
 {4(s-\kappa_1)(s-\kappa_2)\left[\sqrt{(s+\lambda-\kappa_1)^{\alpha}-\lambda^{\alpha}}(\sqrt{(s+\lambda-\kappa_1)^{\alpha}-\lambda^{\alpha}}+\sqrt{(s+\lambda-\kappa_2)^{\alpha}-\lambda^{\alpha}})\right]^3},
\end{split}
\end{equation}
\end{widetext}
where $B_1(s)=2(\alpha-1)\lambda^{\alpha}+(2+\alpha)(s+\lambda-\kappa_1)^{\alpha}$ and $B_2(s)=2(\alpha-1)\lambda^{\alpha}+(2-\alpha)(s+\lambda-\kappa_1)^{\alpha}$. Substitution of $\kappa_1=\kappa_2=0$ into Eq. \eqref{eq:35-2} gives the following special case of our derivation
\begin{equation}\label{35-2-1}
\langle (T^{+})^2\rangle_{s}=\frac{1}{s^3}-\frac{\alpha(s+\lambda)^{\alpha-1}}{4s^2[(s+\lambda)^{\alpha-1}-\lambda^\alpha]},
\end{equation}
which is exactly in agreement with the previous work (Eq. (54) in \cite{Wu2016}).

Particularly, assume $\kappa_1=\kappa_2$, we obtain from Eq. \eqref{eq:35-2} that
\begin{equation}\label{eq:35-3}
  \langle (T^{+})^2\rangle_{s}=\frac{1}{(s-\kappa_1)^3}-\frac{\alpha (s+\lambda-\kappa_1)^{\alpha-1}}{4(s-\kappa_1)^2[(s+\lambda-\kappa_1)^\alpha-\lambda^\alpha]},
\end{equation}
and inversely in the time domain,
\begin{equation}\label{eq:35-4}
\langle (T^{+})^2\rangle=\frac{t^2 e^{\kappa_1 t}}{2}-\frac{\alpha e^{\kappa_1 t}}{4}\int_{0}^{t}(t-\tau)e^{-\lambda \tau}E_{\alpha,1}(\lambda^\alpha \tau^\alpha)d\tau.
\end{equation}
As $t\rightarrow 0$, namely $s\rightarrow \infty$, both $\lambda$ and $\kappa_1$ can be ignored in Eq. \eqref{eq:35-3}. Thus, $\langle (T^{+})^2\rangle \sim \frac{4-\alpha}{8}t^2$, which is the expected result \cite{Wu2016} since initially tempering and reaction terms have negligible influence on the process.

\subsection{First passage time}
\begin{figure}
\centering
\includegraphics[height=4.0cm,width=8.5cm]{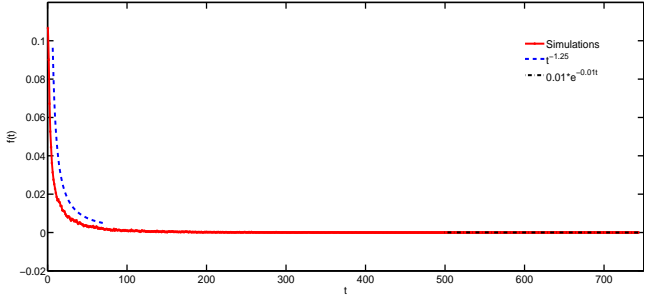}
\caption{PDF $f(t)$ of the first passage time $T_f$ (red) solid line simulated from $10^5$ trajectories when $\kappa_2=-0.01$, $b=5$, $\alpha=0.5$, $\lambda=0$ and $\sigma=1$. The dashed lines are $t^{-1.25}$ (blue) for short but not too short time and $0.01e^{-0.01t}$ (black) for large time scales, respectively. Note that this empirical $f(t)$ decays to zero very fast as $t\rightarrow 0$, which coincides with the theoretical conclusion Eq. (\ref{eq:38}). }\label{fig3}
\end{figure}

\begin{figure}
\centering
\includegraphics[height=4.0cm,width=8.5cm]{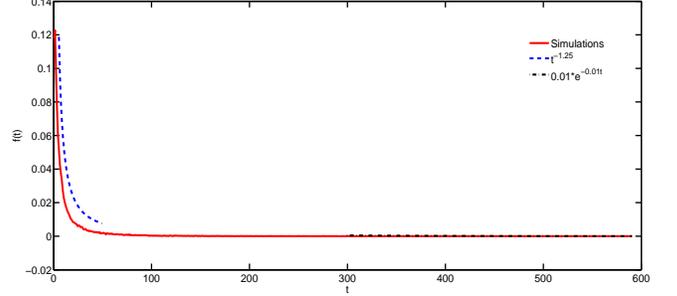}
\caption{PDF $f(t)$ of the first passage time $T_f$ (red) solid line simulated from $5*10^4$ trajectories when $\kappa_2=-0.01$, $b=5$, $\alpha=0.5$, $\lambda=0.5$ and $\sigma=1$. The dashed lines are $t^{-1.25}$ (blue) for short but not too short time and $0.01e^{-0.01t}$ (black) for large time scales, respectively. Note that this empirical $f(t)$ also decays to zero very fast as $t\rightarrow 0$, which coincides with the theoretical conclusion Eq. (\ref{eq:38}). }\label{fig4}
\end{figure}

The time $T_{f}$ when a particle starting at $x_0=-b$ $(b > 0)$ passes $x = 0$ for the first time is called the first passage time \cite{Redner2001}. According to \cite{Kac1951,Wu2016}, the relationship between the distribution
of $T_{f}$ and the PDF of the occupation time in half-space $T^+$ (in the Laplace domain $T^+\rightarrow p $) satisfies
\begin{equation}\label{eq:36}
  P_r\{T_{f}>t\}=\lim_{p \rightarrow\infty}\rho_{-b}(p,t).
\end{equation}
Denote the PDF of $T_{f}$ as $f(t)$, which satisfies $f(t)=-\frac{\partial}{\partial t} \lim_{p \rightarrow\infty}\rho_{-b}(p,t)$ from Eq. \eqref{eq:36}.
According to Eqs. \eqref{eq:33} and \eqref{eq:34},
\begin{equation}\label{eq:37}
\lim_{p \rightarrow\infty}\rho_{-b}(p,s)=\frac{1}{s-\kappa_2}\left(1-e^{-b\sqrt{\frac{(s+\lambda-\kappa_2)^{\alpha}-\lambda^{\alpha}}{K_{\alpha}}}}\right).
\end{equation}
Hence, in the Laplace domain $t\rightarrow s$, we have
\begin{equation}\label{eq:38}
\tilde{f}(s)=\frac{-\kappa_2}{s-\kappa_2}+\frac{s}{s-\kappa_2}e^{-b\sqrt{\frac{(s+\lambda-\kappa_2)^{\alpha}-\lambda^{\alpha}}{K_{\alpha}}}}.
\end{equation}
When $t\rightarrow 0$, corresponding to $s\rightarrow \infty$, we have $\tilde{f}(s)\sim \exp\left(-\frac{b}{ \sqrt{K_\alpha}}s^{\frac{\alpha}{2}}\right)$ from Eq. \eqref{eq:38}, which is equivalent to the one-sided L\'{e}vy law in the time domain. Furthermore, we obtain $\tilde{f}(s)\sim \frac{-\kappa_2}{s-\kappa_2}$ as $s\rightarrow 0$ and equivalently $f(t)\sim -\kappa_2 e^{\kappa_2 t}$ as $t\rightarrow \infty$. Consequently, we assert that $f(t)$ decays very fast to zero when $t\rightarrow 0$, behaves as $t^{-1-\alpha/2}$ for short but not too short time scales and finally decays to zero according to the exponential law $-\kappa_2 e^{\kappa_2 t}$ for adequately large time scales. This conclusion is further confirmed by simulations presented in Figs. \ref{fig3} and \ref{fig4}. Note that the exponential law is distinct from the celebrated $t^{-1.5}$ decay law \cite{Wu2016,Redner2001}, due to the influence of additional reaction terms.

\subsection{Occupation time in half-interval}
In this subsection, we suppose that the particle motion is restricted to the interval $(-L,L)$ with the absorbing or reflecting boundaries. The occupation time $T^{+}_{a}$ or $T^{+}_{r}$ (corresponding to the absorbing or reflecting boundary conditions, respectively) in the positive half-interval is defined as  $T^{+}_{a}=T^{+}_{r}=\int_{0}^{t}U[x(\tau)]d\tau$, where $U(x)=1$ for $0\leq x<L$ and $U(x)=0$ for $-L<x<0$. Thus, we have
\begin{equation}\label{eq:39}
\begin{split}
&\rho_{x_0}(p,s)=
  \\
&\left\{
\begin{array}{ll}
\frac{K_{\alpha}}{(s+\lambda+p-\kappa_1)^{\alpha}-\lambda^{\alpha}}\frac{\partial^2 \rho_{x_0}(p,s)}{\partial x_0^2}+\frac{1}{s+p-\kappa_1},& 0<x_0<L;
\\
\frac{K_{\alpha}}{(s+\lambda-\kappa_2)^{\alpha}-\lambda^{\alpha}}\frac{\partial^2 \rho_{x_0}(p,s)}{\partial x_0^2}+\frac{1}{s-\kappa_2},  &-L<x_0<0.
\end{array}
\right.
\end{split}
\end{equation}

{ (a)} Absorbing boundary conditions

Solving Eq. \eqref{eq:39} in each interval, respectively, we obtain
\begin{widetext}
\begin{equation}\label{eq:40}
  \rho_{x_0}(p,s)= \left\{
  \begin{array}{ll}
  C_1e^{x_0\sqrt{\frac{(s+\lambda+p-\kappa_1)^{\alpha}-\lambda^{\alpha}}{K_{\alpha}}}}+C_2e^{-x_0\sqrt{\frac{(s+\lambda+p-\kappa_1)^{\alpha}-\lambda^{\alpha}}{K_{\alpha}}}}+\frac{1}{s+p-\kappa_1}
  , & 0<x_0<L;
  \\
 C_3e^{x_0\sqrt{\frac{(s+\lambda-\kappa_2)^{\alpha}-\lambda^{\alpha}}{K_{\alpha}}}}+C_4e^{-x_0\sqrt{\frac{(s+\lambda-\kappa_2)^{\alpha}-\lambda^{\alpha}}{K_{\alpha}}}}+\frac{1}{s-\kappa_2}, &-L<x_0<0.
  \end{array}
  \right.
\end{equation}

With the absorbing boundary condition $\rho_{x_0}(p,s)|_{x_0=\pm L=0}$ and the continuity of $\rho_{x_0}(p,s)$ and its first derivative at $x_0=0$, we have
\begin{equation}\label{eq:40-1}
\left\{
\begin{array}{ll}
 2C_1=[1+E(p,s)]C_3+[1-E(p,s)]C_4+\frac{1}{s-\kappa_2}-\frac{1}{s+p-\kappa_1},
 \\
 \\
 2C_2=[1-E(p,s)]C_3+[1+E(p,s)]C_4+\frac{1}{s-\kappa_2}-\frac{1}{s+p-\kappa_1},
 \\
 \\
 C_3=\frac{\frac{1}{s-\kappa_2}\left[(1-e^{LE_2(s)})\cosh LE_1(p,s)-E(p,s)\sinh LE_1(p,s) \right]
  -\frac{1}{s+p-\kappa_1}e^{LE_2(s)}\left[1-\cosh LE_1(p,s)\right]}
 {[1+E(p,s)]\sinh [LE_1(p,s)+LE_2(s)]+
 [1-E(p,s)]\sinh [-LE_1(p,s)+LE_2(s)]},
 \\
 \\
 C_4=\frac{\frac{1}{s-\kappa_2}\left[(-1+e^{-LE_2(s)})\cosh LE_1(p,s)-E(p,s)\sinh LE_1(p,s) \right]
  +\frac{1}{s+p-\kappa_1}e^{-LE_2(s)}\left[1-\cosh LE_1(p,s)\right]}
 {[1+E(p,s)]\sinh [LE_1(p,s)+LE_2(s)]+
 [1-E(p,s)]\sinh [-LE_1(p,s)+LE_2(s)]},
 \end{array}
 \right.
\end{equation}
where we denote for simplicity $E_1(p,s)=\sqrt{\frac{(s+\lambda+p-\kappa_1)^{\alpha}-\lambda^{\alpha}}{K_{\alpha}}}$, $E_2(s)=\sqrt{\frac{(s+\lambda-\kappa_2)^{\alpha}-\lambda^{\alpha}}{K_{\alpha}}}$ and $E(p,s)=\frac{E_2(s)}{E_1(p,s)}$. Suppose the particle departs at $x=0$. Then
\begin{equation}\label{40-2}
  \rho_0(p,s)=\frac{\frac{1}{s-\kappa_2}\left[-1-E(p,s)\frac{\tanh LE_1(p,s)}{\sinh LE_2(s)}\right]-
  \frac{1}{s+p-\kappa_1}\left[\frac{1}{\cosh LE_1(p,s)}-1\right]}
  {1+E(p,s)\frac{\tanh LE_1(p,s)}{\tanh LE_2(s)}}+\frac{1}{s-\kappa_2}.
\end{equation}
\end{widetext}

{ (b)} Reflecting boundary conditions

Solving Eq. \eqref{eq:39} with $\frac{\partial\rho_{x_0}(p,s)}{\partial x_0}|_{x_0=\pm L}=0$ gives
\begin{equation}\label{eq:41}
\small \begin{split}
 & \rho_{x_0}(p,s)=
  \\
  &\left\{
  \begin{array}{ll}
  C_1\cosh\left[(L-x_0)\sqrt{\frac{(s+\lambda+p-\kappa_1)^{\alpha}-\lambda^{\alpha}}{K_{\alpha}}}\right]+\frac{1}{s+p-\kappa_1}, & x_0>0;
  \\
  C_2\cosh\left[(L+x_0)\sqrt{\frac{(s+\lambda-\kappa_2)^{\alpha}-\lambda^{\alpha}}{K_{\alpha}}}\right]+\frac{1}{s-\kappa_2}, & x_0<0.
  \end{array}
  \right.
\end{split}
\end{equation}
Similarly, in order to determine the constants $C_1$ and $C_2$, we assume the continuity of $\rho_{x_0}(p,s)$ and its first derivative at $x_0=0$ in Eq. \eqref{eq:41}. Hence,
\begin{equation}\label{eq:41-1}
\left\{
\begin{array}{ll}
C_1=-F(p,s)C_2,
\\
C_2=\frac{1}{F(p,s)\cosh[LE_1(p,s)]+\cosh[LE_2(s)]}\cdot\frac{\kappa_1-\kappa_2-p}{(s+p-\kappa_1)(s-\kappa_2)}.
\end{array}
\right.
\end{equation}
where we denote $F(p,s)=\frac{E_2(s)\sinh[LE_2(s)]}{E_1(p,s)\sinh[LE_1(p,s)]}$ for simplicity. If the particle starts from $x_0=0$, then
\begin{equation}\label{41-2}
\small
\begin{split}
  \rho_0(p,s)&=\frac{E_1(p,s)\tanh LE_1(p,s)}
  {(s+p-\kappa_1)[\tanh LE_1(p,s)+\tanh LE_2(s)]}
  \\
  &+\frac{E_2(s)\tanh LE_2(s)}
  {(s-\kappa_2)[\tanh LE_1(p,s)+\tanh LE_2(s)]}.
\end{split}
\end{equation}
Especially, when $\lambda=\kappa_1=\kappa_2=0$, we recover
\begin{equation}\label{41-3}
  \rho_0(p,s)=\frac{(s+p)^{\frac{\alpha}{2}-1}\tanh L\sqrt{\frac{(s+p)^\alpha}{K_\alpha}}+s^{\frac{\alpha}{2}-1}\tanh L\sqrt{\frac{s^\alpha}{K_\alpha}}}
  {(s+p)^{\frac{\alpha}{2}}\tanh L\sqrt{\frac{(s+p)^\alpha}{K_\alpha}}+s^{\frac{\alpha}{2}}\tanh L\sqrt{\frac{s^\alpha}{K_\alpha}}},
\end{equation}
which was previously derived in \cite{Carmi2011} using the similar method.

\section{Summary}
The functional distributions of the trajectories of diffusion processes have been well developed, including the ones for normal and anomalous diffusions. Chemical reaction is another important process that results in the transformation of chemical substances. What happens for the functional distribution if we have both reaction and diffusion? This paper is answering this question. The previous ideas of deriving the governing equations for the functional distributions of pure (normal/anomalous) diffusion processes do not work for the general reaction diffusion processes. We provide a theoretical framework of deriving the governing equations for the functional distributions of the trajectories of the stochastic processes with both reaction and diffusion.

It is well known that the net effect of normal diffusion and chemical reaction processes is just the sum of the individual rates of change. However, the net effect becomes complicated for the two processes: anomalous diffusion and chemical reaction. What about the functional distributions of the net effect of the diffusion and reaction processes? We show that for any type of diffusion, if the reactional rate is a constant or only time dependent $r(t)$, then the functional distribution of the diffusion and reaction processes is the one of pure diffusion process multiplied by $e^{\int_{0}^{t}r(u)du}$. For the case with general reaction rates, we derive a series of specific  Feynman-Kac equations with various diffusion types.

%
%
%
%
%

Several functional distributions of interest are investigated by means of analytically solving the derived backward Feynman-Kac equations, including the occupation time in half-space, the first-passage time to a fixed boundary and the occupation time in half-interval with absorbing or reflecting boundary conditions. Compared with the previous work, we are more concerned with the influence of the reaction terms to the temporal-tempered anomalous dynamics and validate some new theoretical conclusions by simulations. It is first found that the mean occupation time in half-space of the reaction diffusion process obeys the $\frac{te^{\kappa_1 t}}{2}$ law, where $\kappa_1$ is the assumed reaction rate in the whole space. The PDF of the first passage time to the fixed boundary $x=0$ starting from $x=-b$ has the asymptotic form $-\kappa_2 e^{\kappa_2 t}$ as $t\rightarrow \infty$, and $\kappa_2$ is the prescribed reaction rate in the negative half-space.


The coming research topics that need to explore should include functional distributions of particle creation processes, the first-passage time to a moving boundary, reaction diffusion dynamics under the influence of external potentials and the corresponding functional distributions.


\section*{Acknowledgments}
This work was supported by the National Natural Science Foundation of China under Grant No.11671182. We thank Eli Barkai for the helpful discussions.

\begin{widetext}
\section*{APPENDIX}
\subsection{Taylor expansion of the first term on the RHS of Eq. \eqref{eq:13}}\label{AppendixA}
We express the second order Taylor expansion of the first term on the RHS of Eq. \eqref{eq:13} for small $z$ as follows:
\begin{equation}\label{eq:A01}
\begin{split}
&~~~~~\int_{\mathbb{R}}\int_{0}^{t}K(t-\tau)\rho(x-z,p,\tau)e^{ip\int_{\tau}^{t}U(x-z,t')dt'}e^{\int_{\tau}^{t}r(\rho(x-z,u))du}w(z)d\tau dz
\\
&\approx\int_{\mathbb{R}}\int_{0}^{t}K(t-\tau)\rho(x,p,\tau)e^{ip\int_{\tau}^{t}U(x,t')dt'}e^{\int_{\tau}^{t}r(\rho(x,u))du}w(z)d\tau dz
\\
&-\int_{\mathbb{R}}\int_{0}^{t}K(t-\tau)\frac{\partial}{\partial x}\left[\rho(x,p,\tau)e^{ip\int_{\tau}^{t}U(x,t')dt'}e^{\int_{\tau}^{t}r(\rho(x,u))du}\right]zw(z)d\tau dz
\\
&+\frac{1}{2}\int_{\mathbb{R}}\int_{0}^{t}K(t-\tau)\frac{\partial^2}{\partial x^2}\left[\rho(x,p,\tau)e^{ip\int_{\tau}^{t}U(x,t')dt'}e^{\int_{\tau}^{t}r(\rho(x,u))du}\right]z^2 w(z)d\tau dz
\\
&=\int_{0}^{t}K(t-\tau)\rho(x,p,\tau)e^{ip\int_{\tau}^{t}U(x,t')dt'}e^{\int_{\tau}^{t}r(\rho(x,u))du}d\tau
\\
&+\frac{\sigma^2}{2}\frac{\partial^2}{\partial x^2}\int_{0}^{t}K(t-\tau)\rho(x,p,\tau)e^{ip\int_{\tau}^{t}U(x,t')dt'}e^{\int_{\tau}^{t}r(\rho(x,u))du}d\tau,
\end{split}
\tag{A1}
\end{equation}
since we have the assumptions $\int_{\mathbb{R}}w(z)dz=1$, $\int_{\mathbb{R}}zw(z)dz=0$ and $\int_{\mathbb{R}}z^2w(z)dz=\sigma^2$.
\end{widetext}
\subsection{Asymptotic forms of the Laplace transforms of (tempered) power-law waiting time PDF}
In the first place, consider the power-law waiting time PDF $\phi(t)=\alpha \tau ^{\alpha} t^{-1-\alpha}\mathbf{1}_{[\tau, +\infty)}(t)$ with $0<\alpha<1$, which obviously satisfies the normalization $\int_{0}^{+\infty}\phi(t)dt=1$. According to the formula $\int_{0}^{+\infty}(e^{iky}-1)\alpha y^{-1-\alpha}dy=-\Gamma(1-\alpha)(-ik)^{\alpha}$ for $0<\alpha<1$ in \cite{Meerschaert2012}, we have
\begin{equation}\label{eq:A02}
\begin{split}
\tilde{\phi}(s)&=\int_{0}^{+\infty}e^{-st}\phi(t)dt=\int_{\tau}^{+\infty}e^{-st}\alpha \tau ^{\alpha} t^{-1-\alpha}dt
\\
&=\int_{\tau}^{+\infty}\alpha \tau ^{\alpha} t^{-1-\alpha}dt+\int_{\tau}^{+\infty}[e^{-st}-1]\alpha \tau ^{\alpha} t^{-1-\alpha}dt
\\
&=1-\tau ^{\alpha}\Gamma (1-\alpha)s ^{\alpha}-\int_{0}^{\tau}[e^{-st}-1]\alpha \tau ^{\alpha} t^{-1-\alpha}dt
\\
&=1-\tau ^{\alpha}\Gamma (1-\alpha)s ^{\alpha}+O(s).
\end{split}
\tag{A2}
\end{equation}
Usually, $O(s)$ is omitted in practice since we are mainly attentive to the long-time asymptotic behaviour of $\phi(t)$, which corresponds to $\tilde{\phi}(s)$ in the limit of $s\rightarrow0$.

Next, we proceed to calculate the Laplace transform of the tempered power-law PDF $\phi(t)=C_{\tau}^{-1}e^{-\lambda t}t^{-1-\alpha}\mathbf{1}_{[\tau, +\infty)}(t)$ ($0<\alpha<1$ and $C_{\tau}$ is the normalization factor) as follows:
\begin{equation}\label{eq:A03}
\begin{split}
 \tilde{\phi}(s)&=\int_{\tau}^{+\infty}(e^{-st}-1+1)C_{\tau}^{-1}e^{-\lambda t}t^{-1-\alpha}dt
 \\
 &=1+\int_{\tau}^{+\infty}(e^{-(s+\lambda)t}-1)C_{\tau}^{-1}t^{-1-\alpha}dt
 \\
 &-\int_{\tau}^{+\infty}(e^{-\lambda t}-1)C_{\tau}^{-1}t^{-1-\alpha}dt
 \\
 &=1-\frac{\Gamma(1-\alpha)}{\alpha C_{\tau}}[(s+\lambda)^{\alpha}-\lambda^{\alpha}]
 \\
 &+\int_{0}^{\tau}(e^{-\lambda t}-e^{-(s+\lambda)t})C_{\tau}^{-1}t^{-1-\alpha}dt
 \\
 &=1-\frac{\Gamma(1-\alpha)}{\alpha C_{\tau}}[(s+\lambda)^{\alpha}-\lambda^{\alpha}]+O(s).
\end{split}
\tag{A3}
\end{equation}
\begin{widetext}
\subsection{Asymptotic forms of the Fourier transform of tempered power-law jump length PDF}
As for the tempered power-law distribution $w(x)=C_{\varepsilon}^{-1}e^{-\gamma |x|}|x|^{-1-\beta}\mathbf{1}_{[\varepsilon, +\infty)}(|x|)$ with $0<\beta<2$ and the normalization factor $C_{\varepsilon}$, we calculate its Fourier transform $\hat{w}(k)=\int_{-\infty}^{+\infty}e^{ikx}w(x)dx$ from two aspects.
First of all, when $0<\beta<1$, we have
\begin{equation}\label{eq:A04}
\begin{split}
 \hat{w}(k)&=1+\int_{\varepsilon}^{+\infty}(e^{ikx}-1)C_{\varepsilon}^{-1}e^{-\gamma x}x^{-1-\beta}dx+\int^{-\varepsilon}_{-\infty}(e^{ikx}-1)C_{\varepsilon}^{-1}e^{-\gamma |x|}|x|^{-1-\beta}dx
 \\
 &=1+\int_0^{+\infty}(e^{(ik-\gamma)x}-1)C_{\varepsilon}^{-1}x^{-1-\beta}dx+\int_0^{+\infty}(e^{-(ik+\gamma)y}-1)C_{\varepsilon}^{-1}y^{-1-\beta}dy-2\int_0^{+\infty}(e^{-\gamma x}-1)C_{\varepsilon}^{-1}x^{-1-\beta}dx
 \\
 &+\int_{0}^{\varepsilon}(e^{-\gamma x}-e^{(ik-\gamma)x})C_{\varepsilon}^{-1}x^{-1-\beta}dx+\int_{0}^{\varepsilon}(e^{-\gamma y}-e^{-(ik+\gamma)y})C_{\varepsilon}^{-1}y^{-1-\beta}dy
 \\
 &=1-\frac{\Gamma(1-\beta)}{\beta C_{\varepsilon}}(\gamma-ik)^{\beta}-\frac{\Gamma(1-\beta)}{\beta C_{\varepsilon}}(\gamma+ik)^{\beta}+\frac{2\Gamma(1-\beta)}{\beta C_{\varepsilon}}\gamma^{\beta}+O(k)
 \\
 &=1-\frac{2\Gamma(1-\beta)\cos\theta\beta}{\beta C_{\varepsilon}}(k^2+\gamma^2)^{\beta/2}+\frac{2\Gamma(1-\beta)}{\beta C_{\varepsilon}}\gamma^{\beta}+O(k),
\end{split}
\tag{A4}
\end{equation}
where $\theta=\arg(\gamma+ik)=\arctan(k/\gamma)$. Thus, as $k\rightarrow 0$, $\cos\theta\beta\rightarrow 1$.
Secondly, when $1<\beta<2$, we have
\begin{equation}\label{eq:A05}
\begin{split}
 \hat{w}(k)&=1+\int_{-\infty}^{+\infty}ik xw(x)dx+\int_{\varepsilon}^{+\infty}(e^{ikx}-ikx-1)C_{\varepsilon}^{-1}e^{-\gamma x}x^{-1-\beta}dx+\int^{-\varepsilon}_{-\infty}(e^{ikx}-ikx-1)C_{\varepsilon}^{-1}e^{-\gamma |x|}|x|^{-1-\beta}dx
 \\
 &=1+\int_0^{+\infty}(e^{(ik-\gamma)x}-(ik-\gamma)x-1)C_{\varepsilon}^{-1}x^{-1-\beta}dx+\int_0^{+\infty}(e^{-(ik+\gamma)y}+(ik+\gamma)y-1)C_{\varepsilon}^{-1}y^{-1-\beta}dy
 \\
 &-2\int_0^{+\infty}(e^{-\gamma x}+\gamma x-1)C_{\varepsilon}^{-1}x^{-1-\beta}dx+\int_{0}^{\varepsilon}(2e^{-\gamma x}-e^{(ik-\gamma)x}-e^{-(ik+\gamma)x})C_{\varepsilon}^{-1}x^{-1-\beta}dx
 \\
 &=1+\frac{\Gamma(2-\beta)}{\beta (\beta-1) C_{\varepsilon}}(\gamma-ik)^{\beta}+\frac{\Gamma(2-\beta)}{\beta(\beta-1) C_{\varepsilon}}(\gamma+ik)^{\beta}-\frac{2\Gamma(2-\beta)}{\beta(\beta-1)C_{\varepsilon}}\gamma^{\beta}+O(k^2)
 \\
 &=1-\frac{2\Gamma(1-\beta)\cos\theta\beta}{\beta C_{\varepsilon}}(k^2+\gamma^2)^{\beta/2}+\frac{2\Gamma(1-\beta)}{\beta C_{\varepsilon}}\gamma^{\beta}+O(k^2),
\end{split}
\tag{A5}
\end{equation}
according to the recurrence relation $\Gamma(1-\beta)=\frac{\Gamma(2-\beta)}{1-\beta}$.
\end{widetext}


\end{document}